\documentclass[aps,pra,amssymb,amsmath,onecolumn
,superscriptaddress,showpacs]{revtex4-1}

\usepackage[latin9]{inputenc}
\usepackage{amsmath}
\usepackage{amssymb}
\usepackage{graphicx}
\usepackage{verbatim}
\usepackage{bm}
\usepackage{color}

\usepackage{graphicx,epsfig,amsfonts,amssymb}
\usepackage{bm}
\usepackage{times}
\usepackage{verbatim}
\usepackage{multirow}
\usepackage{array} 

\newcommand{\nn}{\nonumber}

\newcommand{\ra}{\rangle}
\newcommand{\rar}{\rightarrow}

\newcommand{\be}{\begin{eqnarray}}
\newcommand{\ee}{\end{eqnarray}}

\newcommand{\bs}{\begin{equation}\begin{split}}
\newcommand{\es}{\end{split}\end{equation}}

\date{\today}

\usepackage[colorlinks=true,citecolor=blue,linkcolor=magenta]{hyperref}

\begin{document}

\title{Solution to a class of multistate Landau-Zener model beyond integrability conditions}

\author{Rongyu Hu}
\affiliation{School of Physics and Electronics, Hunan University, Changsha 410082, China}
\author{Fuxiang Li}
\affiliation{School of Physics and Electronics, Hunan University, Changsha 410082, China}
\author{Chen Sun}
\email{chensun@hnu.edu.cn}
\affiliation{School of Physics and Electronics, Hunan University, Changsha 410082, China}

\begin{abstract}
We study a class of multistate Landau-Zener model which cannot be solved by integrability conditions or other standard techniques. By analyzing analytical constraints on its scattering matrix and performing fitting to results from numerical simulations of the Schr\"{o}dinger equation, we find nearly exact analytical expressions of all its transition probabilities for specific parameter choices. We also determine the transition probabilities up to leading orders of series expansions in terms of the inverse sweep rate (namely, in the diabatic limit) for general parameter choices. We further show that this model can describe a Su-Schrieffer-Heeger chain with couplings changing linearly in time. Our work presents a new route, i.e., analytical constraint plus fitting, to analyze those multistate Landau-Zener models which are beyond the applicability of conventional solving methods.
\end{abstract}

\maketitle

\section{Introduction}

Multistate Landau-Zener (LZ) models are quantum models whose Hamiltonians depend on time linearly. They are generalizations of the two-state LZ model \cite{landau,zener,majorana,stuckelberg}  to multistate case, i.e., models with numbers of levels larger than $2$. Unlike the two-state LZ model which is exactly solvable using different methods \cite{Shevchenko-2010,Shevchenko-2023} (solvable in the sense that its transition probabilities for an evolution from $t=-\infty$ to $t=\infty$ can be obtained analytically), a general multistate LZ model cannot be exactly solved. However, at special choices of parameters, exact solvability of a multistate LZ model can be achieved; since the 1960s, different types of such models have been identified \cite{DO,Hioe-1987,bow-tie,GBT-Demkov-2000,GBT-Demkov-2001,chain-2002,4-state-2002,4-state-2015,6-state-2015,Patra-2015,DTCM-2016,DTCM-2016-2,quest-2017,large-class}, including the Demkov-Osherov model \cite{DO}, the bow-tie model \cite{bow-tie}, the generalized bow-tie model \cite{GBT-Demkov-2000,GBT-Demkov-2001}, and the driven Tavis-Cummings model \cite{DTCM-2016,DTCM-2016-2}; they each present classes of solvable Hamiltonians whose number of levels can be arbitrarily large. In 2018, Sinitsyn et al. proposed to define integrability of a time-dependent quantum model by its satisfaction of the so-called ``integrability conditions''  \cite{commute}. They showed that these integrability conditions may be used to find a model's exact solution, providing a unified framework on studying solvable multistate LZ models. Most of the previously discovered solvable multistate LZ models have indeed been proven to satisfy the integrable conditions (note that there are models which are solvable but are yet to be shown to satisfy the integrability conditions, for example, the infinite chain model solved in \cite{chain-2002}) and can indeed be solved by utilizing their integrability, 
and new solvable models have been found in the light of integrability conditions \cite{Yuzbashyan-2018,DSL-2019,MTLZ,parallel-2020,quadratic-2021,nogo-2022}.


However, models that do satisfy the integrability conditions are not necessarily solvable, as pointed out in \cite{quadratic-2021}. On the other hand, exact solvability is possible for certain models without the help of integrability. Besides usage of integrability conditions, the two standard solving methods  are usage of special functions as in Zener's treatment of the two-state LZ model \cite{zener}, and Laplace transformations combined with contour integrations as in Majonara's approach to the two-state LZ model \cite{Wilczek-2014,Kofman-2023}, and in the original works on the Demkov-Osherov model \cite{DO}, the bow-tie model \cite{bow-tie} and the generalized bow-tie model \cite{GBT-Demkov-2001}.  Both methods involve direct treatment of  differential equations. There is also a method that avoids dealing with differential equations and thus may involve simpler calculations compared to the previous two -- let's call it the ``analytical constraint'' method, which makes use of the fact that the scattering matrix of a multistate LZ model satisfies certain analytical constraints. In 1993, Brundolbler and Elser observed that the transition probabilities to stay in the levels with extremal slopes take exact analytical forms for a general multistate LZ model \cite{B-E-1993}. The Brundolbler-Elser formula was later rigorously proved \cite{nogo-2004,Shytov-2004}, which was also extended to a set of constraints on the scattering matrix of a general multistate LZ model (named the hierarchy constraints) \cite{HC-2017}. Besides, the scattering matrix of any models with a Hermitian Hamiltonian is unitary. Moreover, a specific model may possess certain symmetries which generate additional relations among its scattering amplitudes. All these together form a set of analytical constraints on the scattering matrix, and the model is solvable as long as the corresponding set is solvable. Such a method was used in Refs.~\cite{HC-2017}  and \cite{cross-2017} to solve several multistate LZ models (see also Ref.~\cite{Nikolai-2014} on the so-called multistate Landau-Zener-Coulomb models which generalizes multistate LZ models by including terms inversely proportional to time).  The number of unknown scattering amplitudes scales with the number of levels $N$ as $N^2$, whereas the number of constraints generally increases less fast (for example, the numbers of hierarchy and unitarity constraints increase with $N$ linearly), so this analytical constraint method is generally expected to work better for models with small  numbers of levels.

In this work, we consider a type of multistate LZ model which cannot be solved by either integrability conditions or other standard methods. We apply the analytical constraint method described in the previous paragraph to write out a set of constraints on its scattering matrix. We show that although this set is not solvable, the number of remaining independent unknowns is small, and together with results from numerical simulations we are able to express all transition probabilities by analytical functions with good accuracies for particular choices of parameters. Our work thus provides a new route to analyze multistate LZ models not exactly solvable. 

This paper is organized as follows. In Section II, we introduce the multistate LZ model considered, define the scattering problem, and discuss its unsolvability by integrability conditions or other standard methods. In Section III, we analyze constraints on the scattering matrix of this model and show that all its transition probabilities depend on two independent unknowns. In Section IV, we show that, for particular Hamiltonian parameters, these two independent unknowns can be approximately expressed by analytical expressions via fitting to exact numerical simulations. In Section V, we determine the two independent unknowns up to leading orders of series expansions in terms of the inverse sweep rate for general  parameter choices of the Hamiltonian. In Section VI, we show that the considered model can describe a $5$-site Su-Schrieffer-Heeger (SSH) chain under a linear quench. Finally, Section VII presents conclusions and discussions.

\section{The model}

\begin{figure}[!htb]
  (a)\scalebox{0.35}[0.35]{\includegraphics{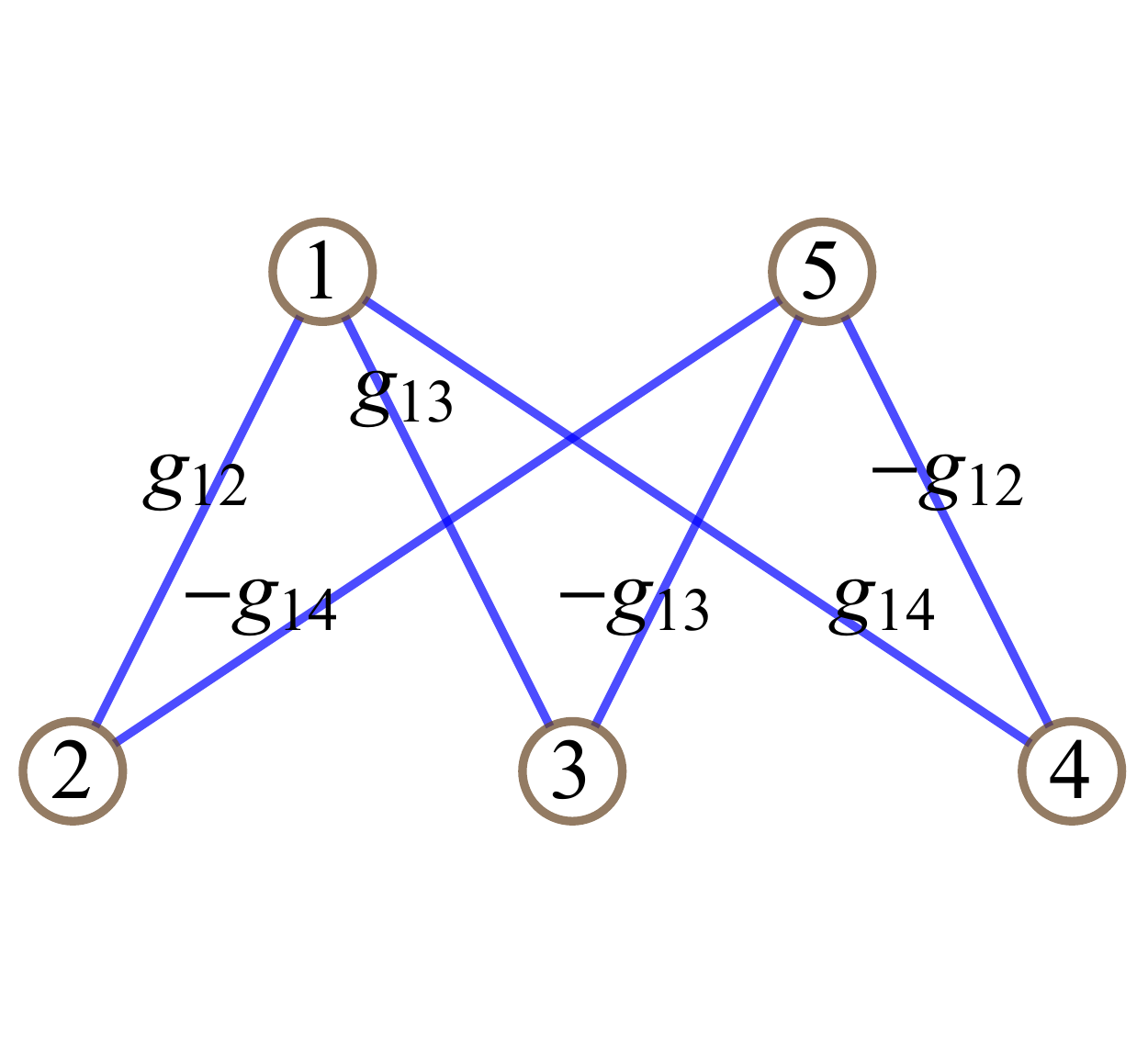}}
    (b)\scalebox{0.4}[0.4]{\includegraphics{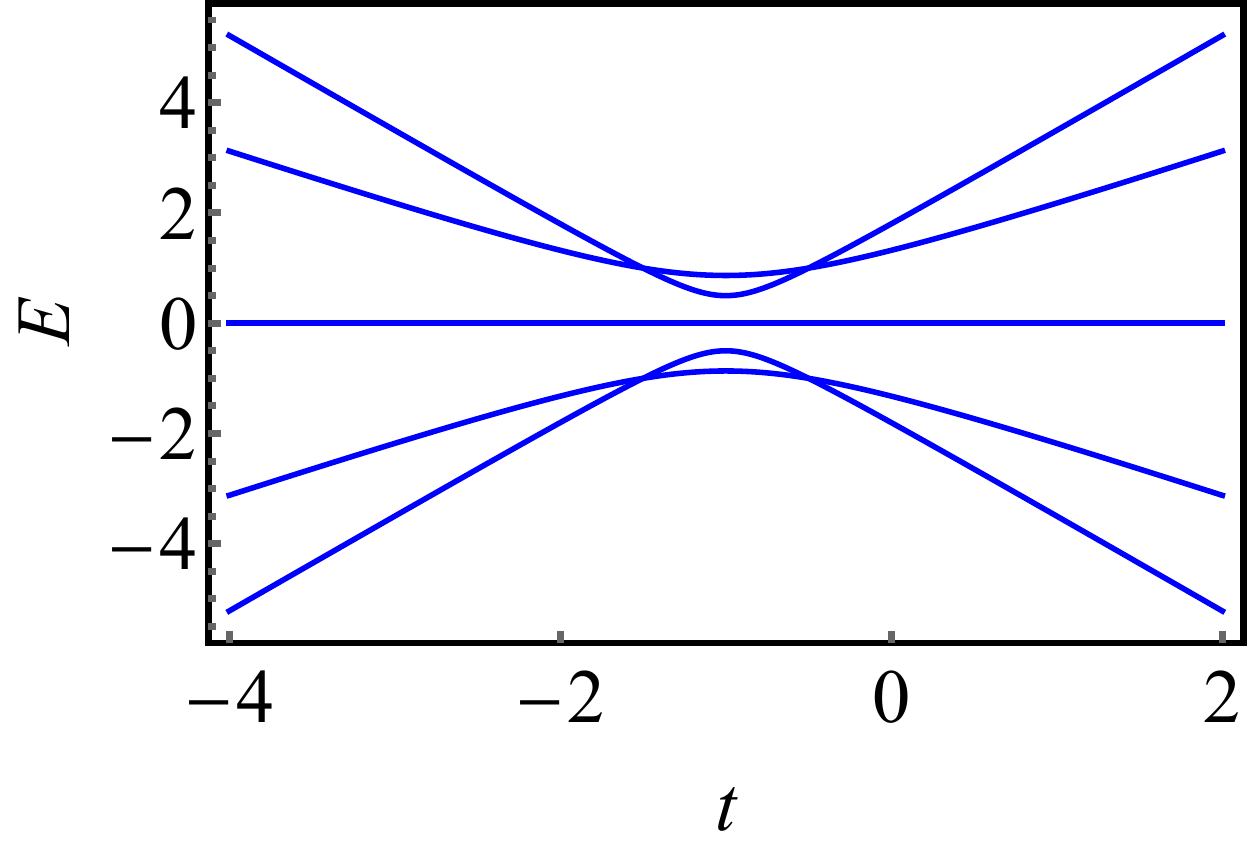}}
      (c)\scalebox{0.4}[0.4]{\includegraphics{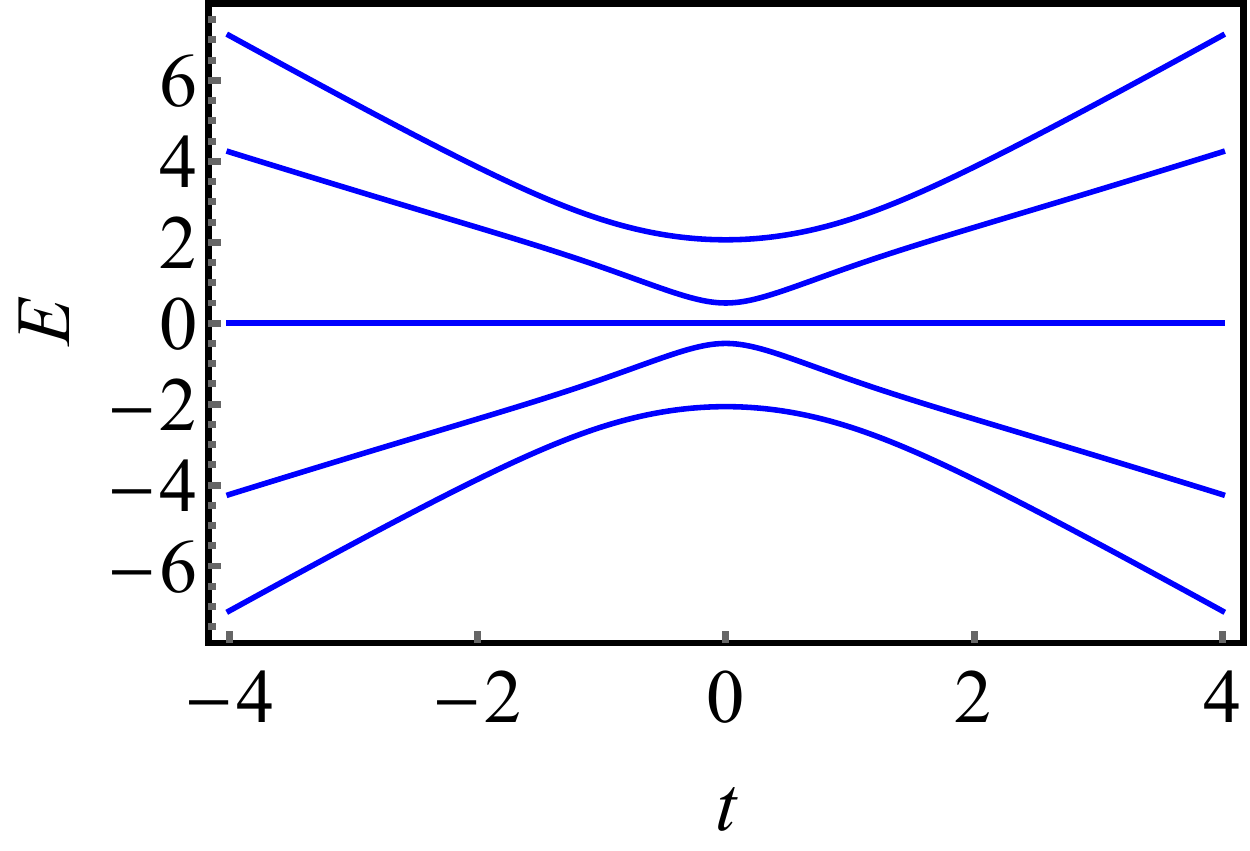}}
\caption{(a) Connectivity graph of the model  \eqref{eq:Schrodinger}, with each vertex corresponding to a diabatic level and each edge to a non-zero coupling between a pair of levels. Each coupling is labelled on its corresponding edge. (b) Adiabatic energy diagram $E$ vs. $t$ of the model  \eqref{eq:Schrodinger} at the parameter choice \eqref{eq:Halpara-SSH} with $\beta=1$. (c) $E$ vs. $t$  of the model  \eqref{eq:Schrodinger} at the parameter choice  \eqref{eq:Halpapa-2nd} with $\beta=1$.}
\label{fig:5-state-graph}
\end{figure}

In this section, we introduce the multistate LZ model considered in this paper. It originates from a consideration of a general odd-size SSH chain with a linear quench of its couplings. This general system can be mapped to a multistate LZ model which seems difficult to solve (these are to be discussed in Section VI). Starting from the cases with smallest number of sites, we find that the simplest yet non-trivial case is a $5$-site chain, and it corresponds to a $5$-state LZ model of the following form:
 \begin{align}\label{}
&i\frac{d\psi}{dt}=H(t) \psi,\label{eq:Schrodinger}\quad H(t)=  \left( \begin{array}{ccccc}
-b_1 t & g_{12}   & g_{13}   &  g_{14}     & 0  \\
g_{12} & -b_2t & 0  &  0  &  -g_{14}\\
g_{13} &  0 & 0 &    0  & -g_{13} \\
g_{14} &  0 &  0  &  b_2 t   & -g_{12}\\
0 & -g_{14}  & -g_{13}  &  -g_{12}  & b_1 t
\end{array} \right),
\end{align}
where the slopes satisfy $b_2>b_1>0$ and the couplings $g_{12}$, $g_{13}$ and $g_{14}$ are all real. This model describes interactions of two groups of levels labelled by $1, 5$ and $2,3,4$, as illustrated in its connectivity graph in Fig.~\ref{fig:5-state-graph}(a). All its diabatic energies (namely, the diagonal elements of the Hamiltonian as functions of time) cross at a single point, and it thus belongs to the class of bipartite models studied in \cite{cross-2017}. Its evolution operator from an initial time $t_i$ to a final time $t_f$ can be formally written as $U(t_f,t_i)={\cal T}e^{-i\int_{t_i}^{t_f} H(t)dt}$, where ${\cal T}$ is a time-ordering operator. We are especially interested in the scattering matrix $S$ of an evolution from $t=-\infty$ to $\infty$, namely, $S=\lim_{T\rar \infty}U(T,-T)$. The transition probability matrix $P$ is connected to $S$ by $P_{ij}=|S_{ij}|^2$. Note that the model \eqref{eq:Schrodinger} is quite general in the sense that all $5$ parameters $b_1$, $b_2$, $g_{12}$, $g_{13}$ and $g_{14}$ can be taken as free independent parameters. In Fig.~\ref{fig:5-state-graph}(b) and (c), the adiabatic energy diagrams (instantaneous eigenvalues of the Hamiltonian as functions of time) of the model \eqref{eq:Schrodinger} are plotted for two specific  sets of parameters \eqref{eq:Halpara-SSH} and \eqref{eq:Halpapa-2nd} in Section IV. In particular, the  parameter choice \eqref{eq:Halpara-SSH} corresponds to the case of a $5$-site SSH model, as will be discussed in Section IV.

One may wonder if the model \eqref{eq:Schrodinger} satisfies the integrable conditions. We first observe that, although it has the same connectivity graph as the $5$-state ``fan'' model which is integrable and exactly solvable \cite{large-class,MTLZ}, it is actually not a fan model, since its order of slopes do not satisfy the constraints for a fan model \cite{large-class}. It thus does not belong to multitime LZ models studied in \cite{MTLZ}, because for a $5$-state LZ model the fan model is the only possibility of a multitime LZ model, as proved in \cite{nogo-2022}. Besides, we do not observe that the model can be constructed as a degenerate limit of a known integrable model with parallel levels. We also considered the Hamiltonian in \eqref{eq:Schrodinger} to depend on another variable $\tau$ and looked for its commutating operator quadratic in time that satisfy integrability conditions following the procedures in \cite{quadratic-2021}. We find that such an operator does exist, but only when we allow the Hamiltonian at a general $\tau$ to violate the chiral symmetry, i.e. to allow the magnitudes of slopes of the states $1$ and $5$ to be different. In other words, the model \eqref{eq:Schrodinger} corresponds to a snapshot --  a special parameter choice which respects the chiral symmetry -- of a more general integrable model. But we do not find this fact helpful in solving the original model, since the underlying integrable model has lower symmetry and seems more difficult to be solved. In sum, the model \eqref{eq:Schrodinger} is indeed integrable in the sense of \cite{commute}, but its integrability seems to bring no help in solving it.

Attempts of solving the model  \eqref{eq:Schrodinger} by other standard methods are also not successful. The special function method is somehow designed for two-state cases -- it relies on reducing the original two first order differential equations to a single second order one whose solutions are known special functions, and it does not work for the current $5$-state model. The Laplace transformation method may work in the situation that after the transformation one obtains a system of equations with at most one differential equation and with all the rest being algebraic equations, but in our case we obtain a system of equations containing two differential equations, and the system seems not solvable. We also point out that another method to find approximate closed-form solutions to time-dependent quantum models, named the exact WKB method \cite{Aoki-2002,Shimada-2020}, is also not applicable here. This method assumes pairwise crossings of diabatic energy levels, but the model \eqref{eq:Schrodinger} has a simultaneous crossing of $5$ diabatic levels.

The analytical constraint method seems to be the last hope to obtain a solution of  \eqref{eq:Schrodinger} in analytical forms; but note that numerically solving the Schr\"{o}dinger equation is also a commonly-used method to treat any quantum problem \cite{Ashhab-2016}, and numerical solutions may provide useful information on the expressions of transition probabilities. In fact, a combination of these two methods allows us to find nearly analytical solutions to \eqref{eq:Schrodinger} in a wide range of parameters, as will be discussed in the next two sections.

\section{Constraints on the scattering matrix}

In this section, we attack the model \eqref{eq:Schrodinger} by the analytical constraint method outlined in the second paragraph of Section I. We first notice that the model \eqref{eq:Schrodinger} possesses certain symmetries. 
Its Schr\"{o}dinger equation  is invariant under both symmetry operations below (let's write $\psi=(a_1,a_2,a_3,a_4,a_5)^T$):\\

\smallskip
\noindent 1. (Time reversal) $t\rar -t$, and $a_1\rar-a_1$, $a_5\rar-a_5$;\\
2. (Chiral) $H\rar -H$, and $a_1\leftrightarrow a_5$, $a_2\leftrightarrow a_4$.\\
\smallskip

These two symmetries can be equivalently expressed by the following operations on the Hamiltonian:
\begin{align}\label{}
&\textrm{Time reversal:} \quad H(-t)=-\Theta H(t) \Theta ,\label{eq:time-reversal}\\
&\textrm{Chiral:} \quad H(t)=-\Gamma H(t) \Gamma,\label{eq:chiral}
\end{align}
where $\Theta=\operatorname{diag}(-1,1,1,1,-1)$  is a diagonal matrix, and $\Gamma$ is an anti-diagonal matrix with all anti-diagonal elements equal to $1$.
The time-reversal symmetry is a general property of any bipartite model as discussed in \cite{cross-2017} (see Eq. (9) in  \cite{cross-2017}), so the considered model \eqref{eq:Schrodinger}, being within the class of bipartite models, satisfy it as well. The chiral symmetry is a result of the special arrangement of the slopes and couplings in the Hamiltonian. It dictates the Hamiltonian's adiabatic energy diagram ($E$ vs. $t$) to be symmetric under a reflection about the $E=0$ line.

The symmetry operations on the evolution matrix $U(t_f,t_i)$ corresponding to \eqref{eq:time-reversal} and \eqref{eq:chiral} read:
\begin{align}\label{}
&\Theta U(t_f,t_i) \Theta
=U^\dag(-t_i,-t_f),\\
&\Gamma U(t_f,t_i) \Gamma=U^*(t_f,t_i),
\end{align}
respectively. Taking $-t_i=t_f=T$, we find that the scattering matrix satisfies $\Theta S \Theta= S^\dag$ and $\Gamma S \Gamma= S^*$. The most general form of $S$ satisfying these two relations is:
\begin{align}\label{}
&S=\left( \begin{array}{ccccc}
S_{11} & -S_{21}^*  &  -S_{31}^*  &   -S_{41}^*    &  S_{51}^* \\
S_{21} & S_{22}  &  S_{32}^*  &   S_{42}^*    &  S_{41}^* \\
S_{31} & S_{32}  &  S_{33}  &   S_{32}^*    &  S_{31}^* \\
S_{41} & S_{42}  &  S_{32}  &   S_{22}    &  S_{21}^* \\
S_{51} & -S_{41}  &  -S_{31}  &   -S_{21}    &  S_{11}
\end{array} \right),
\label{eq:S-symmetry}
\end{align}
with $S_{11}$, $S_{22}$ and $S_{33}$ real. This scattering matrix contains $6$ complex unknowns and $3$ real unknowns. Effectively, there are $6\times2 +3=15$ real unknowns. To find solutions of the scattering problem, one needs to reduce the number of independent unknowns in $S$ as much as possible, which we will do next.

First, the time-reversal symmetry leads to one constraint among the diagonal elements of $S$ as derived in \cite{cross-2017} for a general bipartite model, which for our model reads:
\begin{align}\label{eq:cons-bipartite}
2S_{22}+S_{33} -2S_{11} =1.
\end{align}
Second, the matrix $S$ has to be unitary since the Hamiltonian is Hermitian. This gives $9$ more relations, $3$ of which involve only magnitudes:
\begin{align}\label{}
&S_{11}^2+|S_{21}|^2+|S_{31}|^2+|S_{41}|^2+|S_{51}|^2=1,\label{eq:unitary-1}\\
&|S_{21}|^2+S_{22}^2+|S_{32}|^2+|S_{42}|^2+|S_{41}|^2=1,\\
&2|S_{31}|^2+2|S_{32}|^2+S_{33}^2=1,\label{eq:unitary-3}
\end{align}
and the rest involve also phases:
\begin{align}
&2S_{11}S_{51}+2S_{21}S_{41}+S_{31}^2=0,\label{eq:unitary-4}\\
&2S_{21}^*S_{41}+2S_{22}S_{42}+S_{32}^2=0,\\
&-S_{11}S_{21}+S_{21}S_{22}+S_{31}S_{32}^*+S_{41}S_{42}^*-S_{51}S_{41}^*=0,\\
&-S_{11}S_{31}+S_{21}S_{32}+S_{31}S_{33}+S_{41}S_{32}^*-S_{51}S_{31}^*=0,\\
&-S_{11}S_{41}+S_{21}S_{42}+S_{31}S_{32}+S_{41}S_{22}-S_{51}S_{21}^*=0,\\
&S_{21}^* S_{31}+S_{22}S_{32}+ S_{32}S_{33}+S_{42}S_{32}^*+S_{41}S_{31}^*=0.\label{eq:unitary-9}
\end{align}
Note that the symmetry-required form \eqref{eq:S-symmetry} of $S$ and all the constraints above apply to an evolution $U(T,-T)$ for any $T$, not necessarily at the limit $T\rar \infty$. When this limit is taken, there exists another type of constraints. Namely, it was found in  \cite{HC-2017} that the scattering matrix for a $-\infty$ to $\infty$ evolution of any multistate LZ model satisfies the so-called hierarchy constraints (HCs). For the model \eqref{eq:Schrodinger}, they read:
\begin{align}\label{}
&S_{22}=X,\label{eq:cons-HC1}\\
&\det \left( \begin{array}{cc}
S_{11} & -S_{21}^*      \\
S_{21} & S_{22}     \\
\end{array} \right)=Y,\label{eq:cons-HC2}\\
&\det \left( \begin{array}{c cc}
S_{11} & -S_{21}^*  &  -S_{31}^*     \\
S_{21} & S_{22}  &  S_{32}^*   \\
S_{31} & S_{32}  &  S_{33}    \\
\end{array} \right)=Y,\label{eq:cons-HC3}\\
&\det \left( \begin{array}{cccc}
S_{11} & -S_{21}^*  &  -S_{31}^*  &         S_{51}^* \\
S_{21} & S_{22}  &  S_{32}^*  &     S_{41}^* \\
S_{31} & S_{32}  &  S_{33}  &    S_{31}^* \\
S_{51} & -S_{41}  &  -S_{31}  &     S_{11}
\end{array} \right)=X,\label{eq:cons-HC4}
\end{align}
where
\begin{align}\label{}
X=\sqrt{p_{12}p_{14} },\quad Y=\sqrt{p_{13}} p_{14},
\end{align}
with
\begin{align}\label{}
&p_{12}=e^{-2\pi \Omega_{12}},\quad  \Omega_{12}=\frac{g_{12}^2}{b_2-b_1},\label{eq:Omega12}\\
&p_{13}=e^{-2\pi \Omega_{13}},\quad \Omega_{13}=\frac{g_{13}^2}{b_1},\label{eq:Omega13}\\
&p_{14}=e^{-2\pi \Omega_{14}},\quad \Omega_{14}=\frac{g_{14}^2}{b_1+b_2}\label{eq:Omega14}.
\end{align}
Equations \eqref{eq:cons-bipartite}-\eqref{eq:cons-HC4} are all the constraints we found on the elements of $S$ in the form \eqref{eq:S-symmetry}. To see whether this system of equations can be solved, we need to compare the number of unknowns and that of independent equations.

At first sight, the number of real unknowns seems to be $15$. But a detailed analysis, which we present in Appendix A, shows that this system of equations actually does not depend on $6$ independent phases -- only $4$ combinations of phases appear in it. Let's define $4$ ``rotated'' elements as:
\begin{align}
&\tilde{S}_{21}= S_{21}  e^{i(-\arg S_{31}+\arg S_{32})},\label{eq:S21-tilde}\\
&\tilde{S}_{41}= S_{41} e^{i(-\arg S_{31}-\arg S_{32})},\\
&\tilde{S}_{51}= S_{51} e^{i(- 2 \arg S_{31}) } , \\
&\tilde{S}_{42}= S_{42} e^{i(-2 \arg S_{32} ) }.\label{eq:S42-tilde}
\end{align}
Replacing $S_{21}$, $S_{41}$, $S_{51}$ and $S_{42}$ by their corresponding rotated elements, one finds that the phases of $ S_{31} $ and $ S_{32}$ drop out from all the equations. Namely, the constraints \eqref{eq:cons-bipartite}-\eqref{eq:cons-HC4} can be expressed by $5$ real variables $S_{11}$, $S_{22}$, $S_{33}$, $|S_{31}|$ and $|S_{32}|$, and $4$ complex variables $\tilde{S}_{21}$, $\tilde{S}_{41}$, $\tilde{S}_{51}$ and $\tilde{S}_{42}$. Therefore, the number of real unknowns is $13$ instead of $15$.

On the other hand, our analysis also show that only $11$ of the $14$ equations are independent (see Appendix A). Therefore, there are $2$ more unknowns than independent equations. Thus, these equations cannot determine all the unknowns, but they reduce the number of independent unknowns to $2$. One can express all other unknowns in terms of two appropriately chosen ones. For example, if we choose  $S_{33}$ and $|S_{32}|$ as independent, then all other real elements and magnitudes of complex elements are expressed as:
\begin{align}
&S_{11}=\frac{1}{2}(S_{33}-1)+X,\label{eq:S11}\\
&S_{22}=X,\\
&|S_{21}|^2=-\frac{X}{2} (S_{33}-1)-X^2+Y,\\
&|S_{31}|^2=\frac{1}{2}(1-S_{33}^2)-|S_{32}|^2,\\
&|S_{41}|^2   =-\frac{X}{2} (S_{33}-1)-X^2+Z,\\
&|S_{51}|^2 
=\frac{1}{4}(1+S_{33})^2+ |S_{32}|^2 +X^2-Y- Z,\\
&|S_{42}|^2 
=1 + X(S_{33}-1)- |S_{32}|^2+X^2-Y-Z,\label{eq:S42}
\end{align}
where we defined the combination
\begin{align}\label{eq:Z}
Z=\frac{[X(1+S_{33})- |S_{32}|^2]^2}{4Y}
\end{align}
to simplify the expressions. The $4$ rotated elements can also be determined in terms of $S_{33}$ and  $|S_{32}|$ up to an overall sign of their phases, as shown in Appendix A. Thus, from the set of constraints on $S$, all unknowns can be determined in terms of two of them, up to a sign of the phases.

Note that this does not mean that the scattering matrix $S$ depends only on two unknowns. Recall that the phases of $S_{31}$ and $S_{32}$ do not enter these set of equations; they also become independent unknowns. Thus, the total number of independent unknowns in $S$ is $4$. Actually by considering the limiting behavior at $|t|\rar\infty$, one see that $S_{31}$ and $S_{32}$ depend on $t$ as $e^{i[b_1 t^2/2+O(t)]}$ and $e^{i[b_2 t^2/2+O(t)]}$, respectively. These time-dependencies of course are not captured by the above constraints on the scattering matrix. But these complications on phases do not matter as long as only transition probabilities are concerned. From Eqs.~\eqref{eq:S11}-\eqref{eq:S42}, we conclude that by using constraints on the scattering matrix there are two independent transition probabilities.


\section{Solution at specific parameter choices by fitting}

The analytical constraint analysis above shows that it is not possible to determine all the unknowns from the constraints -- two of the unknowns, e.g. $S_{33}$ and $|S_{32}|$, need to be determined by other means. 
Thus, we will seek a solution by numerically solving the Schr\"{o}dinger equation of the model \eqref{eq:Schrodinger} and {\it fitting} the two independent unknowns by analytical expressions. Note that the analytical constraint analysis applies to an arbitrary choice of the set of parameters $b_1$, $b_2$, $g_{12}$, $g_{13}$ and $g_{14}$. But to perform a fitting, we will let a single parameter sweep across a range and fix all the rest parameters. We will choose $b_1$ and $b_2$ to be both proportional to be a sweeping parameter $\beta$, which characterizes the rate of time dependence of the Hamiltonian.

As shown in this section, by this ``analytical constraint plus fitting'' method one is able to find nearly analytical solution to the model \eqref{eq:Schrodinger} at a general set of parameters when the ratio $b_2/b_1$ is not large (essentially below $b_2/b_1= 3\sqrt3$).

\subsection{Solution at parameters corresponding to an SSH chain}
Let's first consider the Hamiltonian in \eqref{eq:Schrodinger} with the following parameter choice:
\begin{align}\label{}
&b_1=\beta,\quad b_2=\sqrt{3}\beta,\quad g_{12}=-\frac{1}{12}(3-\sqrt{3}),\quad g_{13}=-\frac{\sqrt{3}}{3},\quad g_{14}=\frac{1}{12}(3+\sqrt{3}).\label{eq:Halpara-SSH}
\end{align}
Such a (somehow absurd-looking) set of parameters corresponds to a 5-site SSH model, which we will discuss in the section after next.

\begin{figure}[!htb]
(a)\scalebox{0.62}[0.62]{\includegraphics{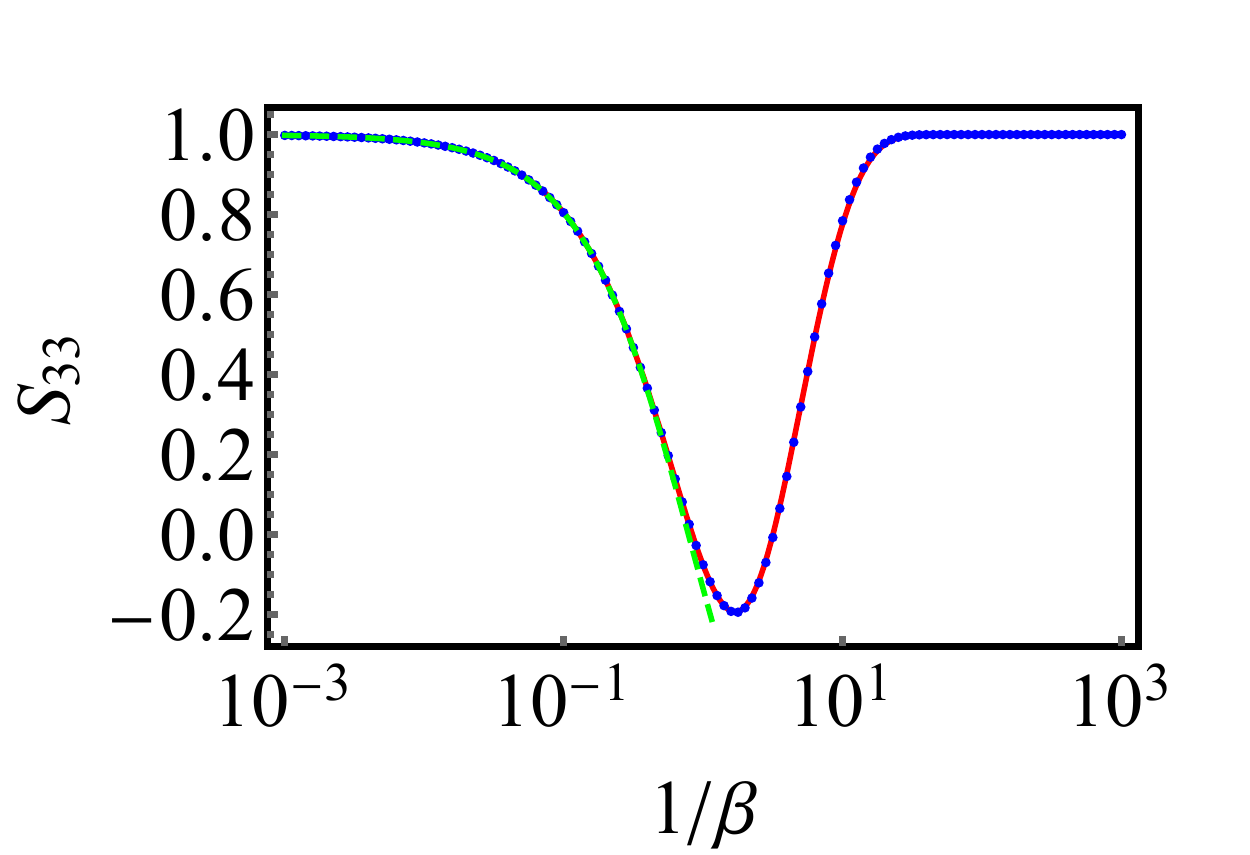}}
(b)\scalebox{0.6}[0.6]{\includegraphics{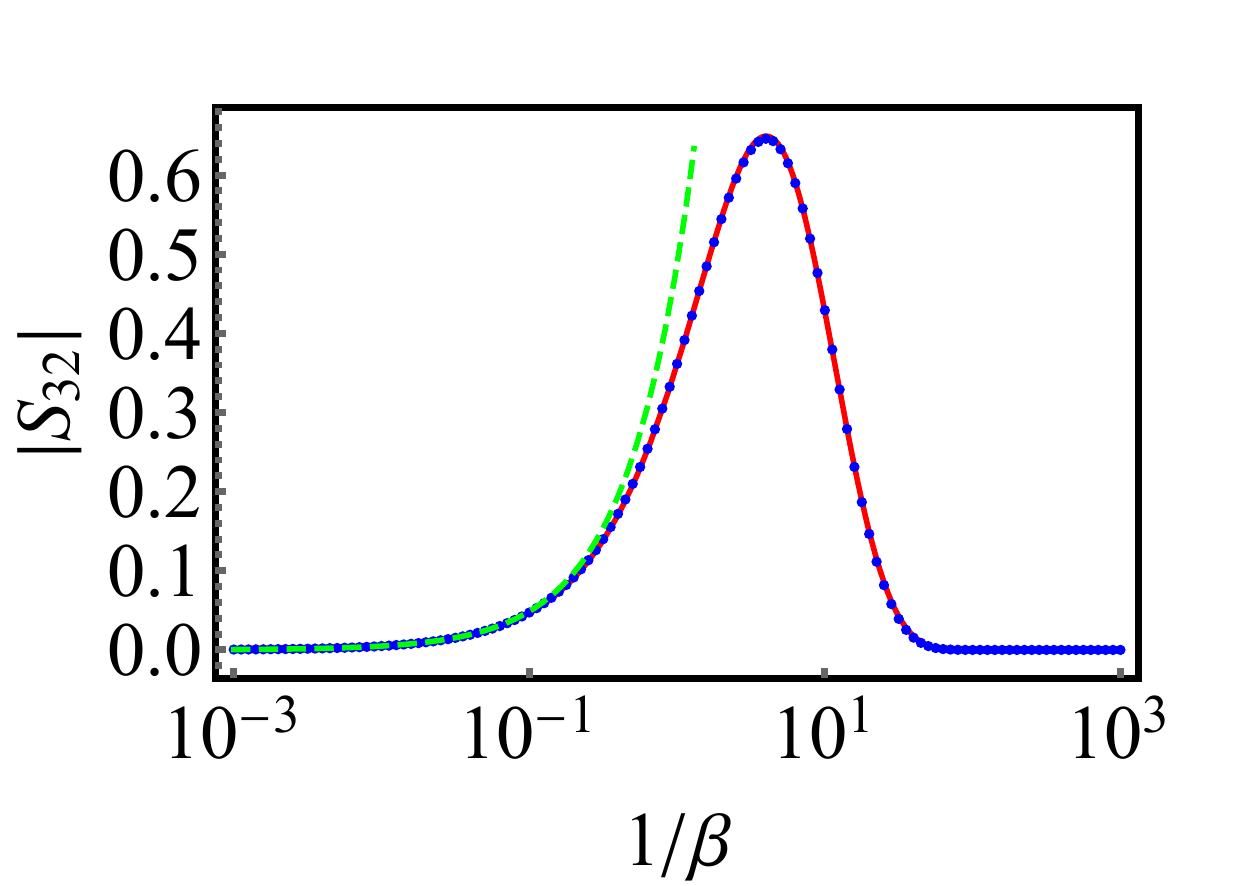}}\\
(c)\scalebox{0.65}[0.65]{\includegraphics{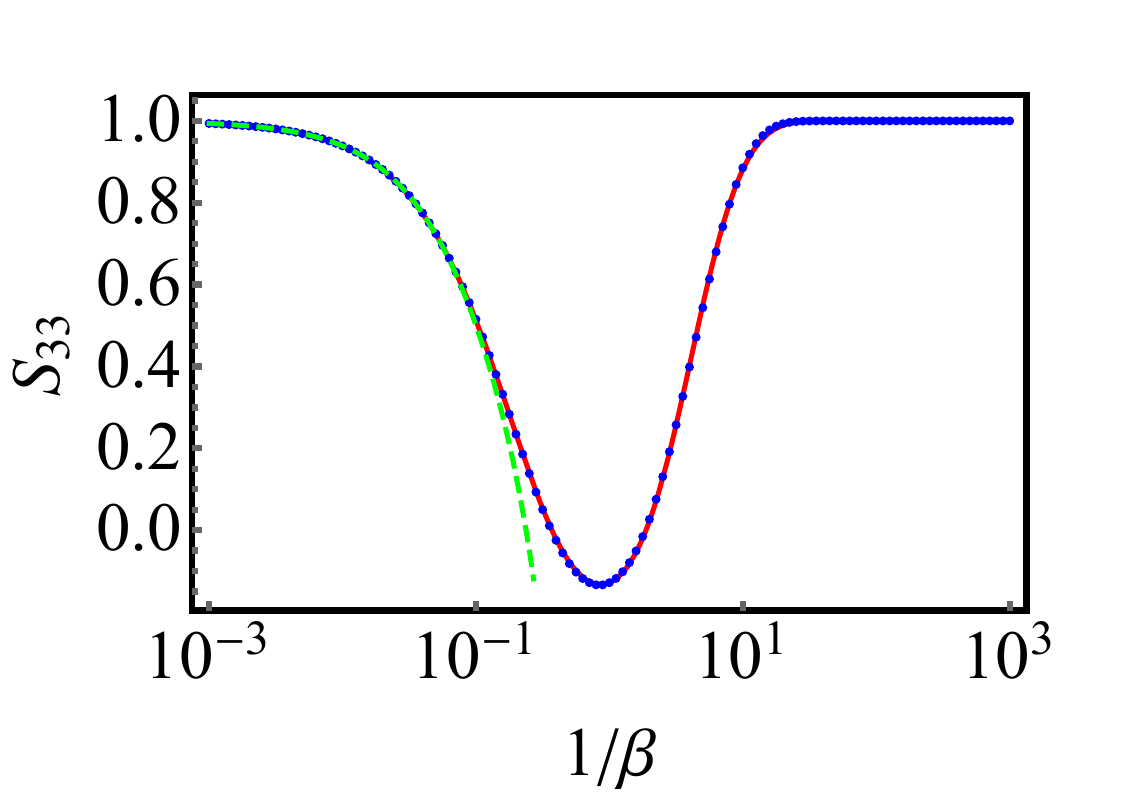}} ~  ~
(d)\scalebox{0.66}[0.66]{\includegraphics{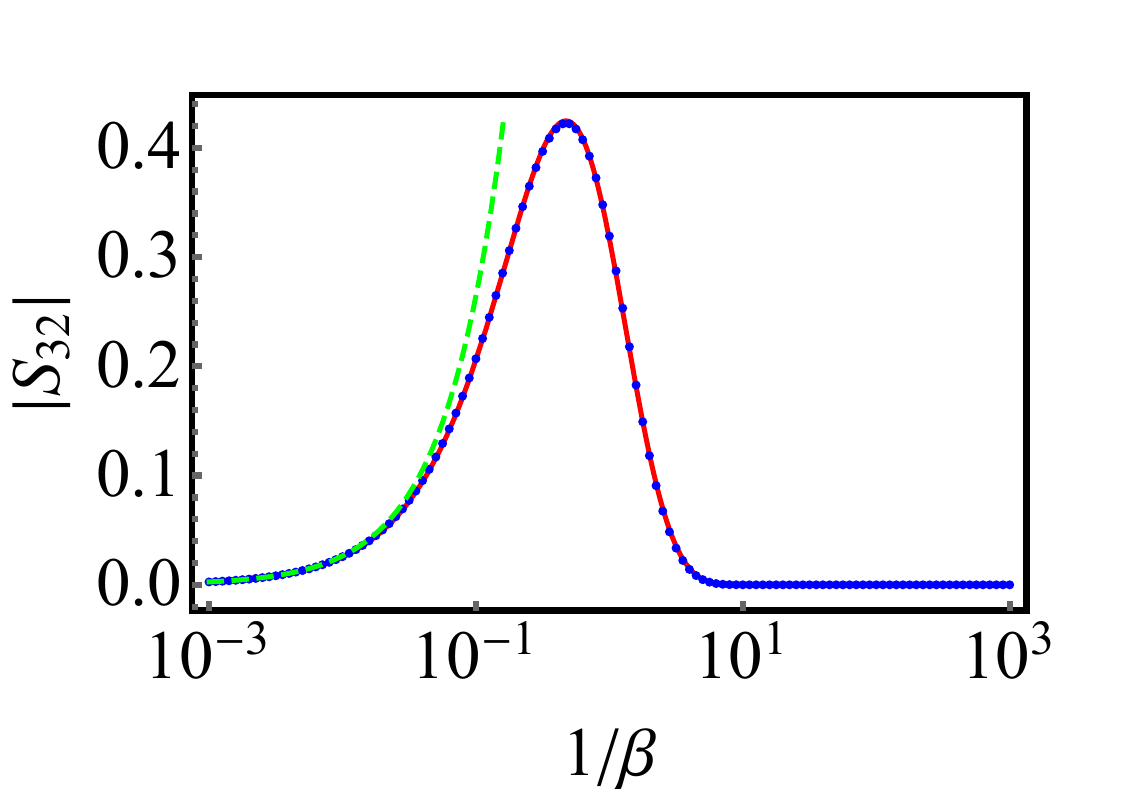}}
\caption{The numerically calculated $S_{33}$ and $|S_{32}|$ vs. $1/\beta$ (blue dots), their corresponding fittings (red solid curves), and series expansions (green dashed curves) for evolution under the Hamiltonian in \eqref{eq:Schrodinger} at two different parameter choices: (a) and (b) at \eqref{eq:Halpara-SSH}, (c) and (d) at \eqref{eq:Halpapa-2nd}. 
The fitting functions are: (a) $S_{33}$ fitted by \eqref{eq:S33fit} with parameters \eqref{eq:S33fitpara}; (b) $|S_{32}|$ fitted by \eqref{eq:S32fit} with parameters \eqref{eq:S32fitpara}; (c) $S_{33}$ fitted by \eqref{eq:S33fit2} with parameters  \eqref{eq:S33fitpara-2nd}; (d) $|S_{32}|$ fitted by \eqref{eq:S32fit} with parameters \eqref{eq:S32fitpara-2nd}. 
The Series expansions are from \eqref{eq:S32series} and \eqref{eq:S33series}.}
\label{fig:S33S32_fit}
\end{figure}

We performed numerical simulations of the Schr\"{o}dinger equation with the Hamiltonian in \eqref{eq:Schrodinger} with its parameters given by Eq.~\eqref{eq:Halpara-SSH} across a wide range of $\beta$. The evolution time is from $t=-T$ to $t=T$, with $T$ taken large enough so that LZ transitions happen in a time much smaller than $T$. In Fig.~\ref{fig:S33S32_fit}(a) and (b), we plot the numerical results of $S_{33}$ and $|S_{32}|$ vs. the inverse sweeping rate $1/\beta$ as dots, with $10^{-3}\le \beta \le 10^3$. 
After some tries, we find that both of them can be fitted quite accurately by simple analytical expressions involving exponentials of the form $e^{-const./\beta}$ (such a form appears commonly in scattering amplitudes of solvable multistate LZ models):
\begin{align}
& S_{33}\approx(1 - 2 e^{-\frac{u_1}{\beta}})(1 - 2 e^{-\frac{v_1}{\beta}}),\label{eq:S33fit}\\
&  |S_{32}|\approx  w e^{-\frac{u_2}{\beta}} (1-e^{-\frac{v_2}{\beta}}),\label{eq:S32fit}
\end{align}
where the fitting parameters (keeping $3$ digits) are given by
\begin{align}\label{}
&u_1=0.219,\quad v_1=0.827,\label{eq:S33fitpara}\\
&u_2=0.108,\quad v_2=0.373,\quad w=1.29.\label{eq:S32fitpara}
\end{align}
These fittings are shown in Fig.~\ref{fig:S33S32_fit}(a) and (b) as solid curves. 
Since other $7$ real elements or magnitudes of complex elements are related to them by \eqref{eq:S11}-\eqref{eq:S42}, we have obtained analytical expressions of all the magnitudes the scattering amplitudes, and thus  analytical expressions of all the transition probabilities.

We note that the fittings \eqref{eq:S33fit} and \eqref{eq:S32fit}, although quite accurate, are not exact. Within the range plotted in Fig.~\ref{fig:S33S32_fit}(a) and (b), the maximal difference between each fitting and its corresponding exact numerical result has an absolute value of roughly $5\times 10^{-3}$. 
Thus, the two analytical expressions are good approximations to the true scattering elements instead of their exact expressions. For this reason, we call these solutions {\it nearly} exact. One may improve the accuracy of fittings at the expense of more fitting parameters. But that results in more complicated analytical expressions, and this procedure could be endless unless one can make a very clever guess of the analytical expressions of the true scattering elements. In fact, it could well be that the true scattering elements are not in forms of any known analytical expressions made of elementary or special functions (if this is the case, the solution of the scattering problem itself in a sense {\it defines} new special functions). Thus, we will not seek more accurate fittings beyond  \eqref{eq:S33fit} and \eqref{eq:S32fit}. 

\subsection{Solutions at other parameter choices}

The fitting functions \eqref{eq:S33fit} and \eqref{eq:S32fit} with fitting parameters \eqref{eq:S33fitpara} and \eqref{eq:S32fitpara} are for the specific parameter choice \eqref{eq:Halpara-SSH} of the Hamiltonian in \eqref{eq:Schrodinger}. One may wonder at a different choice of Hamiltonian's parameters does the fitting method still work, and if so can this be achieved by the same fitting functions  \eqref{eq:S33fit} and \eqref{eq:S32fit}. To answer this question, we performed numerical calculations at  different parameter choices. 
For example, consider a parameter choice:
\begin{align}\label{}
&b_1=\beta,\quad b_2=\sqrt{3}\beta,\quad g_{12}= \frac{1}{2},\quad g_{13}=g_{14}=1,\label{eq:Halpapa-2nd}
\end{align}
where the slope are the same as \eqref{eq:Halpara-SSH}, but the couplings are  quite different. We find that for $|S_{32}|$ the previous fitting function \eqref{eq:S32fit} still work, with fitting parameters:
\begin{align}\label{}
&u_2=1.04,\quad v_2=2.70,\quad w=0.964.\label{eq:S32fitpara-2nd}
\end{align}
For $S_{33}$, the previous fitting function \eqref{eq:S33fit} produces noticeable deviation from the numerics, but we find that a fitting function generalized from \eqref{eq:S33fit} works well:
\begin{align}
& S_{33}\approx(1 - e^{-\frac{u_{1,1}}{\beta}}- e^{-\frac{u_{1,2}}{\beta}})(1 - e^{-\frac{v_{1,1}}{\beta}}-e^{-\frac{v_{1,2}}{\beta}}),\label{eq:S33fit2}
\end{align}
with the $4$ fitting parameters being
\begin{align}\label{}
&u_{1,1}=0.226,\quad u_{1,2}=0.563,\quad v_{1,1}=0.438,\quad v_{1,2}=5.12.\label{eq:S33fitpara-2nd}
\end{align}
These fittings are shown in Fig.~\ref{fig:S33S32_fit}(c) and (d). The previous function \eqref{eq:S33fit} is reduced from \eqref{eq:S33fit2} by requiring $u_{1,1}=u_{1,2}$ and $v_{1,1}=v_{1,2}$. 
At another parameter choice with a larger $b_2/b_1$:
\begin{align}\label{}
&b_1=\beta,\quad b_2= 3\beta,\quad g_{12}= \frac{1}{2},\quad g_{13}=g_{14}=1,\label{}
\end{align}
we find that \eqref{eq:S32fit}  and  \eqref{eq:S33fit2} still works, with parameters $u_{1,1}=0.138$, $u_{1,2}=0.236$, $v_{1,1}=0.803$, $v_{1,2}=5.28$, and $u_2=0.476$, $v_2=2.89$, $w=0.660$. We also identify special cases where the fit function completely fails, but for expected reasons. Consider a choice where all three couplings are identical:
\begin{align}\label{}
&b_1=\beta,\quad b_2=\sqrt{3}\beta,\quad g_{12}=  g_{13}=g_{14}=1.\label{eq:Halpapa-3nd}
\end{align}
At this choice, although $|S_{32}|$ can still be fitted well by the same function \eqref{eq:S32fit}, $S_{33}$ now has a completely  different shape compared to  \eqref{eq:S33fit2}. In fact, at the parameter choice \eqref{eq:Halpapa-3nd} the model experiences a degeneracy -- calculation of eigenvalues of the Hamiltonian shows that at $g_{12}=g_{14}$, instantaneous eigenvalues of the Hamiltonian at $t=0$ experiences an exact crossing point of three levels, whereas this crossing is avoided at any $g_{12}\ne g_{14}$. 
This affects the asymptotic value of $S_{33}$ as $\beta\rar 0$, changing it from $1$ to $-1$. As a result, the previous fitting function \eqref{eq:S33fit2} certainly should not work at such this parameter choice. 

We checked roughly $20$ sets of the Hamiltonian's parameters scattered in the parameter space with no degeneracies. We find that at all choices with  a not very large ratio $b_2/b_1$ (below $b_2/b_1=3\sqrt{3}$), $S_{33}$ and $|S_{32}|$ can be fitted well by the functions \eqref{eq:S33fit2} and \eqref{eq:S32fit}, respectively. But when $b_2/b_1$ becomes large, deviations of the fittings from the exact numerical curves become visible. 
Therefore, the fitting functions  \eqref{eq:S32fit} and \eqref{eq:S33fit2}  work in the region of the parameter space of the Hamiltonian in \eqref{eq:Schrodinger} when $b_2/b_1$ is not large, excluding those points with degeneracies, e.g. at $g_{12}=g_{14}$. In this region, the fitting parameters 
should depend on parameters of the Hamiltonian. A natural question then arises: is it possible to express these dependencies explicitly by analytical functions?
Unfortunately, we do not arrive at definite expressions of these dependencies -- our tries of some simple functions do not work. It's possible that these expressions have complicated forms, and may require considerably more numerical data to be finally revealed. 


\section{Series expansions in the diabatic limit}

In the fitting method above the fitting parameters are not related to the parameters of the original Hamiltonian, which certainly limits the usage of the fitting solutions. In this section, we find expressions of the leading orders of series expansions of the scattering amplitudes $S_{11}$ and $|S_{32}|$  in terms of the inverse scattering rate $1/\beta$, which present  asymptotically exact solutions in the diabatic limit. 
These expressions depends directly on the Hamiltonian's parameters and on no fitting parameters.


The diabatic limit is reached at the sweeping rate $\beta\rar \infty$, or at  $1/\beta\rar 0$. This physically means that the time-dependence is too fast so that no transition between diabatic states can take place. To the zeroth order of $1/\beta$, the scattering matrix is simply a $5\times 5$ unit matrix, and specifically $S_{11}=1+O(1/\beta)$ and $|S_{32}|=0+O(1/\beta)$. We are going to consider these two scattering amplitudes of higher order in their series expansions in $1/\beta$. For this purpose it is useful to notice special cases of parameters when the model is exactly solvable. To present dependencies on $\beta$ explicitly, we define
\begin{align}\label{}
\gamma_{i}/\beta\equiv \pi \Omega_{1i },\quad i=2,3,4,
\end{align}
where $\Omega_{1i }$ was defined in \eqref{eq:Omega12}-\eqref{eq:Omega14}. We identify two exactly solvable cases:

1) $g_{13}=0$. In this case, the state $3$ decouples from all other states, so the scattering amplitude $S_{33}$ is directly given by $S_{33}=1$, and $|S_{32}|=0$. According to the analytical constraints, one thus find that $S_{11}=X=e^{-(\gamma_{2}+\gamma_{4})/\beta}$, and all the rest scattering amplitudes can be determined.

2) $g_{12}=g_{14}=0$. In this case, either states $2$ or state $4$ does not couple to any other states, so $|S_{32}|=0$. The reduced Hamiltonian for the states $1$, $3$ and $5$ form a $3$-state bow-tie model which is exactly solvable \cite{bow-tie} and gives $S_{11}=e^{-\gamma_{3}/\beta}$. Again, all the rest scattering amplitudes can be determined.

As we turn on $g_{12}$, $g_{13}$ and $g_{14}$, generally the scattering amplitudes are expected to deviate continuously from their values in this limits. 
This is confirmed by numerics. Thus, these exact solutions provide starting points for solutions at more general choices of parameters. 


We now consider $S_{11}$ in the diabatic limit. We know from the two exactly solvable cases that:
\begin{align}\label{}
&S_{11}(g_{13}=0)=e^{-(\gamma_{2}+\gamma_{4})/\beta},\\
&S_{11}(g_{12}=g_{14}=0)=e^{- \gamma_{3}/\beta}.
\end{align}
A trial expression for $S_{11}$ that reduces to both of these special cases is $e^{- (\gamma_{2}+\gamma_{3}+\gamma_{4})/\beta}$. From numerical calculations at different values of parameters, we indeed find that this ansatz correctly produces $S_{11}$ up to first order in $1/\beta$. 
At second order in $1/\beta$, numerics reveal an additional contribution of the form $-\gamma_2\gamma_3/\beta^2$ that is universal for different parameter choices. Namely, $S_{11}$ can be determined up to second order in $1/\beta$:
\begin{align}\label{}
&S_{11}  =e^{-(\gamma_{2}+\gamma_{3}+\gamma_{4})/\beta}-\gamma_2\gamma_3/\beta^2+ O(1/\beta^3)\nn\\
&=1-(\gamma_{2}+\gamma_{3}+\gamma_{4})/\beta + (\gamma_{2}^2+\gamma_{3}^2+\gamma_{4}^2+2\gamma_2\gamma_4+2\gamma_3\gamma_4) /(2\beta^2)+ O(1/\beta^3).
\end{align}
Series expansion to a higher order faces difficulties. Our numerics indicate that  the third order term depend not only on  $\gamma_{2}/\beta$, $\gamma_{3}/\beta$ and $\gamma_{4}/\beta$, but also on the ratio $b_2/b_1$. With the current numerical data, this dependence seems to be too complicated to decipher.


For $|S_{32}|$, the two exactly solvable cases both give $|S_{32}|=0$, so there is no additional information other than the zeroth-order general result in diabatic limit. Numerics at different parameter sets show that the first order term has a universal form that depends only on  $\gamma_{2}/\beta$, $\gamma_{3}/\beta$ and $\gamma_{4}/\beta$:
\begin{align}\label{eq:S32series}
&|S_{32}|  =\sqrt{\gamma_3(\gamma_2+\gamma_4)}/\beta+ O(1/\beta^2).
\end{align}
Again, numerics indicate that  the second order term depends also on the ratio $b_2/b_1$, and its determination seems difficult.

Note that instead of $S_{33}$ as in the previous section, we now worked with $S_{11}$ since it turns out to be more convenient. The two are  related simply by \eqref{eq:S11}, and we get the series expansion of $S_{33}$:
\begin{align}\label{eq:S33series}
&S_{33}=1+2[ e^{-(\gamma_{2}+\gamma_{4})/\beta}(e^{-\gamma_{3}/\beta}-1)-\gamma_2\gamma_3/\beta^2]+ O(1/\beta^3).
\end{align}

These series expansions work for all parameter sets we evaluated, and thus they should be universal for the model \eqref{eq:Schrodinger}. For example, for the two parameter sets \eqref{eq:Halpara-SSH} and \eqref{eq:Halpapa-2nd} they are shown as green dashed curves in Fig.~\ref{fig:S33S32_fit}. One can see that the series expansion curves mimic the exact numerical solutions up to roughly $1/\beta\sim10^{-1}$ and deviate at larger $1/\beta$.




\section{Connection to a $5$-site SSH chain}

The SSH model is a famous model on hopping of electrons along a diatomic chain \cite{SSH1,SSH2}. In this section, we are going to show that the $5$-state model \eqref{eq:Schrodinger} with a choice of the set of parameters in \eqref{eq:Halpara-SSH} physically corresponds to a $5$-site SSH chain which couplings change linearly in time.

\subsection{Mapping of an odd-sized SSH chain under a linear quench to a multistate LZ model}
We will first consider a general odd-sized SSH chain with $2N-1$ sites with zero onsite energies, whose Hamiltonian reads:
\begin{align}\label{}
H_{SSH}=\left( \begin{array}{ccccccc}
 0 & J_1 & 0 & 0  & \cdots & 0 & 0 \\
  J_1 & 0 & J_2 & 0 & \cdots & 0 & 0\\
 0 & J_2 & 0 & J_1  & \cdots & 0 & 0\\
  0 &  0 &  J_1 & 0  & \cdots & 0 & 0\\
\vdots&\vdots&\vdots&\vdots&\ddots&\vdots&\vdots\\
0 & 0 & 0 & 0 & \cdots &0 & J_2\\
0 & 0 &0 &  0 & \cdots & J_2 & 0
\end{array} \right).
\label{eq:SSH-Hamiltonian}
\end{align}
We take the couplings to change linearly in time as:
\begin{align}\label{eq:SSH-quench}
J_1=\frac{1}{2}+\beta t,\quad J_2=\frac{1}{2}-\beta t.
\end{align}
We will call it a linear quench. At $t=0$ the two couplings equal, and at $t=\pm 1/(2\beta)$ the chain is fully dimerized.

The Hamiltonian \eqref{eq:SSH-Hamiltonian} depends linearly on time. 
We will transform it to the so-called diabatic basis which is the standard form of a multistate LZ model, which has its time-dependent part resting only on the diagonal elements. For this purpose, we first rewrite \eqref{eq:SSH-Hamiltonian} as
\begin{align}\label{}
H_{SSH}=Bt+A,
\end{align}
where
\begin{align}
B=\beta\left( \begin{array}{ccccccc}
 0 & 1 & 0 & 0  & \cdots & 0 & 0 \\
  1 & 0 & -1 & 0 & \cdots & 0 & 0\\
 0 & -1 & 0 & 1  & \cdots & 0 & 0\\
  0 &  0 &  1 & 0  & \cdots & 0 & 0\\
\vdots&\vdots&\vdots&\vdots&\ddots&\vdots&\vdots\\
0 & 0 & 0 & 0 & \cdots &0 & -1\\
0 & 0 &0 &  0 & \cdots & -1 & 0
\end{array} \right),
\label{}
\end{align}
and
\begin{align}\label{}
A=\frac{1}{2}\left( \begin{array}{ccccccc}
 0 & 1 & 0 & 0  & \cdots & 0 & 0 \\
 1 & 0 & 1 & 0 & \cdots & 0 & 0\\
 0 & 1 & 0 & 1  & \cdots & 0 & 0\\
  0 &  0 &  1 & 0  & \cdots & 0 & 0\\
\vdots&\vdots&\vdots&\vdots&\ddots&\vdots&\vdots\\
0 & 0 & 0 & 0 & \cdots &0 & 1\\
0 & 0 &0 &  0 & \cdots & 1 & 0
\end{array} \right).
\end{align}
A transformation to the diabatic basis is then realized by a unitary transformation to diagonalize of the matrix $B$. The eigenvalues of $B$ read:
\begin{align}\label{}
&\epsilon_j=2\sin\frac{k_j}{2},
\end{align}
where $k_j=\pi j/N$ with $j=- N+1,\ldots,0,\ldots ,N-1$. The corresponding eigenstates are:
\begin{align}\label{}
&\psi_j=(\psi_{1,j},\psi_{2,j},\ldots,\psi_{2N-1,j})^T,\\
&\psi_{x,j}=\frac{1}{\sqrt{N}}\sin(\frac{1}{2}k_j x), \quad \textrm{for } x \textrm{ even,}\nn \\
&\psi_{x,j}=\frac{1}{\sqrt{N}}\cos(\frac{1}{2}k_j x), \quad \textrm{for } x \textrm{ odd,}\nn
\end{align}
where $x=1,2,\ldots, 2N-1$ labels the sites. A unitary transformation with the matrix $V=(\psi_1,\psi_2,\ldots,\psi_{2N-1})$ then transforms $H$ to the diabatic basis, which we denote as $H_{LZ}$:
\begin{align}\label{eq:H-LZ}
&H_{LZ}=V^T H_{SSH} V= V^T B V t+V^T A V\equiv B_{LZ}t+A_{LZ}.
\end{align}
Since all elements of the matrix $V$ are real, this transformation is also orthogonal. The matrix $B_{LZ}$ is, by construction, a diagonal matrix made of the eigenvalues:
\begin{align}\label{eq:BLZ}
&(B_{LZ})_{l,l}=2\beta\sin (\frac{1}{2}k_{l}),
\end{align}
where $l=-N+1,\ldots,N-1$ (for convenience, we labelled the  rows and columns of the matrix $B_{LZ}$ from $-N+1$ to $N-1$ instead of from $1$ to $2N-1$). The elements of the matrix $A_{LZ}$ is more difficult to obtain, but we find that they can also be written in closed forms -- evaluation of $V^T A V$ show that they can be presented as imaginary parts of sums of certain geometric sequences, performing which leads to the result
\begin{align}\label{eq:ALZ}
&(A_{LZ})_{l,m}=\frac{1}{N} \frac{\cos\frac{k_{l}}{2}\cos\frac{k_{m}}{2}}{\sin\frac{k_{l}}{2}+\sin\frac{k_{m}}{2}},\quad \textrm{for odd $m+l$},\nn\\
&(A_{LZ})_{l,m}=0, \quad \textrm{otherwise},
\end{align}
where $l=-N+1,\ldots,N-1$ and $m=-N+1,\ldots,N-1$. Therefore, an odd-sized SSH chain \eqref{eq:SSH-Hamiltonian} under a linear quench \eqref{eq:SSH-quench} can be mapped to a multistate LZ Hamiltonian in the standard form \eqref{eq:H-LZ} whose elements are in simple closed forms.

Now that the Hamiltonian $H_{LZ}=B_{LZ}t+A_{LZ}$ is of the standard form of an multistate LZ model, we can read out several of its properties. First, since $A_{LZ}$'s diagonal elements are zero, the diabatic energy levels all cross at the same point. Second, the states are separated into two groups: one group for states with even indices and the other group for states with odd indices, and the coupling between any two states within the same group is zero. It thus belongs to the class of bipartite models studied in \cite{cross-2017}. Finally, as a result of the chiral symmetry of the original SSH model, the adiabatic energy diagram of $H_{LZ}$ is symmetric under a reflection of $E\rar -E$, and there is a state with permanent zero energy -- the state with index $0$.

Regardless of the above properties, the general multistate LZ model \eqref{eq:H-LZ} at an arbitrary $N$ is very complicated, and it is not obvious at all if it is solvable. Nevertheless,  the very existence of this mapping from an SSH chain to a multistate LZ Hamiltonian with relatively simple expressions of matrix elements is interesting by itself, and we hope our work can stimulate further studies on the model \eqref{eq:H-LZ} with matrix elements \eqref{eq:BLZ} and \eqref{eq:ALZ}. Hopefully analytical solution of this general model can finally be revealed, possibly by more advanced methods developed in the future. At a small $N$ the model  \eqref{eq:H-LZ} is simple enough though. At $N=2$, the model can be recognized as a $3$-state bow-tie model \cite{bow-tie}, which is indeed integrable and exactly solvable. At $N=3$, the Hamiltonian is a $5\times 5$ matrix. Calculating out the matrix elements using the above expressions of $(B_{LZ})_{l,l}$ and $(A_{LZ})_{l,m}$, one finds that this Hamiltonian is just the previously discussed Hamiltonian in \eqref{eq:Schrodinger} with its parameters given by Eq.~\eqref{eq:Halpara-SSH}, if the levels labelled by $ -2,-1,0,1,2$ in \eqref{eq:H-LZ} are identified as levels $2,1,3,5,4$ in \eqref{eq:Schrodinger}, respectively. Thus, our result of the solution of the model \eqref{eq:Schrodinger} can be applied in studying a $5$-site SSH chain under a linear quench, which we discuss in the next subsection.

\subsection{$5$-site SSH chain under a linear quench}
\begin{figure}[!htb]
(a)\scalebox{0.45}[0.45]{\includegraphics{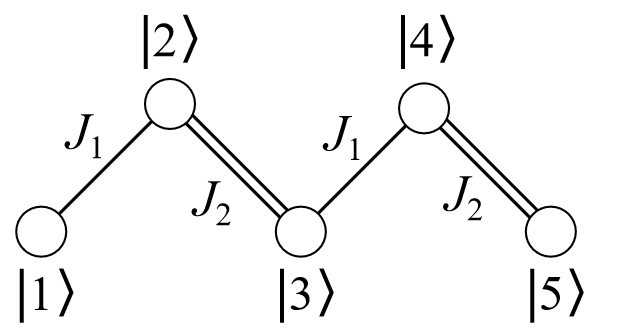}} ~ ~ ~ ~ ~
(b)\scalebox{0.35}[0.35]{\includegraphics{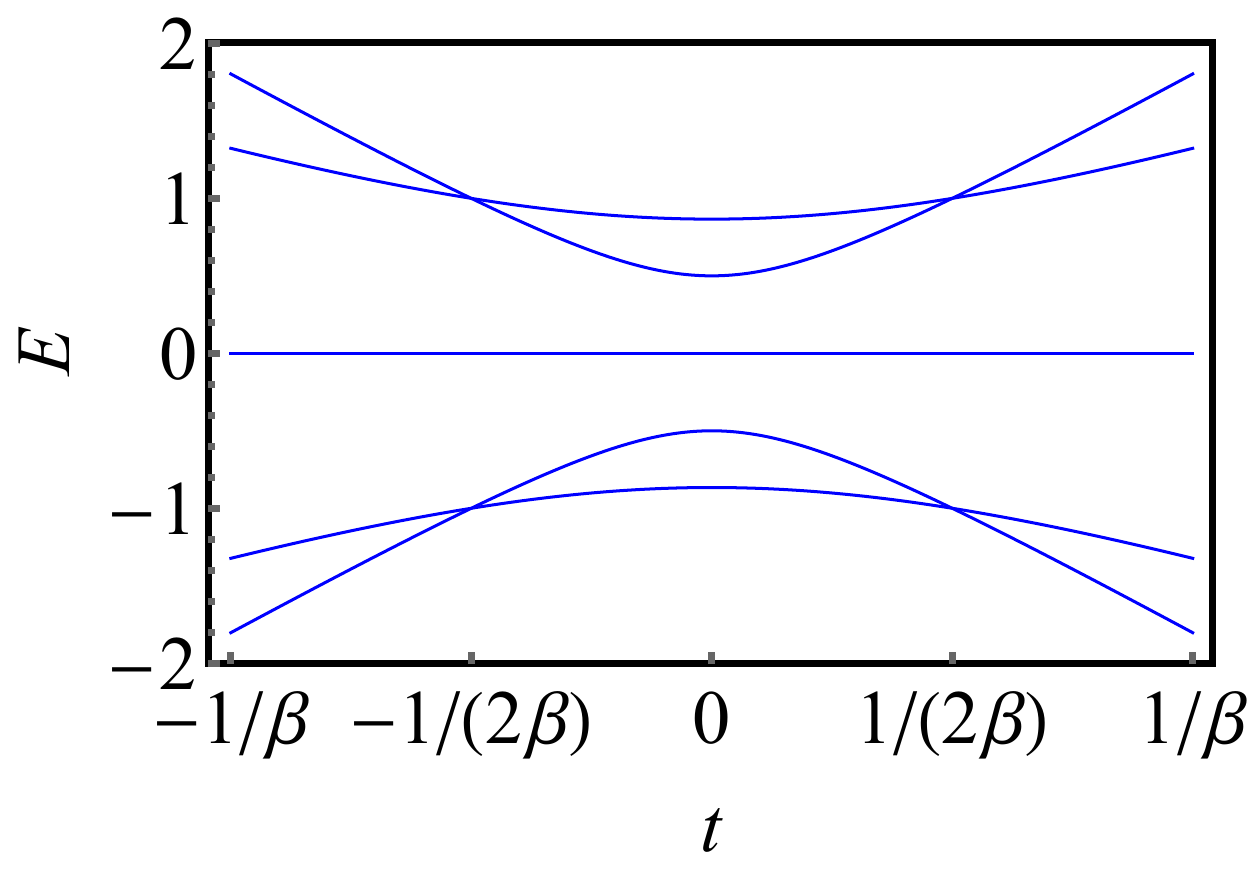}}\\
\bigskip
(c)\scalebox{0.45}[0.45]{\includegraphics{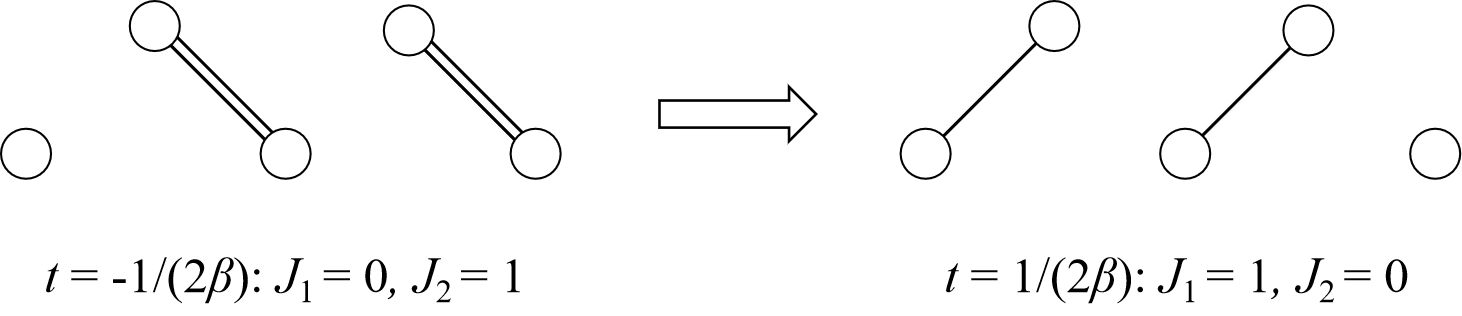}}
\caption{A 5-state SSH chain under a linear quench: (a) a sketch of the chain with $5$ sites; (b) the adiabatic energy diagram $E$ vs. $t$; (c) The chain under a quench from $t=-1/(2\beta)$ to $t=1/(2\beta)$ that connects the two dimerized limits.}
\label{fig:5-SSH}
\end{figure}

The Hamiltonian of a $5$-site SSH chain under a linear quench reads:
\begin{align}\label{}
H_{5,SSH}=\left( \begin{array}{ccccc}
 0 & J_1 & 0 & 0  & 0 \\
  J_1 & 0 & J_2 & 0 & 0\\
 0 & J_2 & 0 & J_1   & 0\\
 0 & 0 & J_1 &0 & J_2\\
0 &  0 & 0 & J_2 & 0
\end{array} \right).
\label{eq:5-SSH-Hamiltonian}
\end{align}
where $J_1=\frac{1}{2}+\beta t$ and $J_2=\frac{1}{2}-\beta t$ as before. An illustration of this model is shown in Fig.~\ref{fig:5-SSH}(a). Fig.~\ref{fig:5-SSH}(b) shows the adiabatic energy diagram, containing a state with permanent zero energy -- the topological edge state. Especially, at two special times $t=\pm 1/(2\beta)$ the couplings take $J_1=0$ or $J_2=0$, respectively, i.e. the chain splits into isolate dimers and an isolate site at the left/right end. A quench from $t=- 1/(2\beta)$ to $t= 1/(2\beta)$ connects the two dimerized limits, as illustrated in Fig.~\ref{fig:5-SSH}(c).

As discussed in the last subsection, the Hamiltonian \eqref{eq:5-SSH-Hamiltonian} is connected to the Hamiltonian in \eqref{eq:Schrodinger} with parameters \eqref{eq:Halpara-SSH} by an orthogonal transformation:
\begin{align}\label{}
&H_{5,LZ}=V^T H_{5,SSH} V.
\end{align}
The evolution operator of the SSH chain $U_{5,SSH}(t_f,t_i)$ is then related to that of the multistate LZ model by the same transformation:
\begin{align}\label{}
&U_{5,SSH}(t_f,t_i)={\cal T}e^{-i\int_{t_i}^{t_f} H_{5,SSH}dt}\nn\\
&={\cal T}e^{-i\int_{t_i}^{t_f} V H_{5,LZ} V^T dt}= V {\cal T}e^{-i\int_{t_i}^{t_f}H_{5,LZ} dt} V^T\nn\\
&=V  U_{5,LZ}(t_f,t_i)V^T.
\end{align}
For this $5$-site chain we have $N=3$, and the explicit form of the matrix $V$ is:
\begin{align}\label{}
V=\left(
\begin{array}{ccccc}
 \frac{1}{2 \sqrt{3}} & \frac{1}{2} & \frac{1}{\sqrt{3}} & \frac{1}{2} & \frac{1}{2 \sqrt{3}} \\
 -\frac{1}{2} & -\frac{1}{2} & 0 & \frac{1}{2} & \frac{1}{2} \\
 -\frac{1}{\sqrt{3}} & 0 & \frac{1}{\sqrt{3}} & 0 & -\frac{1}{\sqrt{3}} \\
 \frac{1}{2} & -\frac{1}{2} & 0 & \frac{1}{2} & -\frac{1}{2} \\
 \frac{1}{2 \sqrt{3}} & -\frac{1}{2} & \frac{1}{\sqrt{3}} & -\frac{1}{2} & \frac{1}{2 \sqrt{3}} \\
\end{array}
\right).
\end{align}
Thus, we can express any scattering amplitude of the $5$-site SSH chain in terms of those of the multistate LZ model \eqref{eq:Schrodinger}. For an evolution from $t=-T$ to $t=T$ with a large $T$, the results in the previous two sections can then be used. 

\begin{figure}[!htb]
\scalebox{0.6}[0.6]{\includegraphics{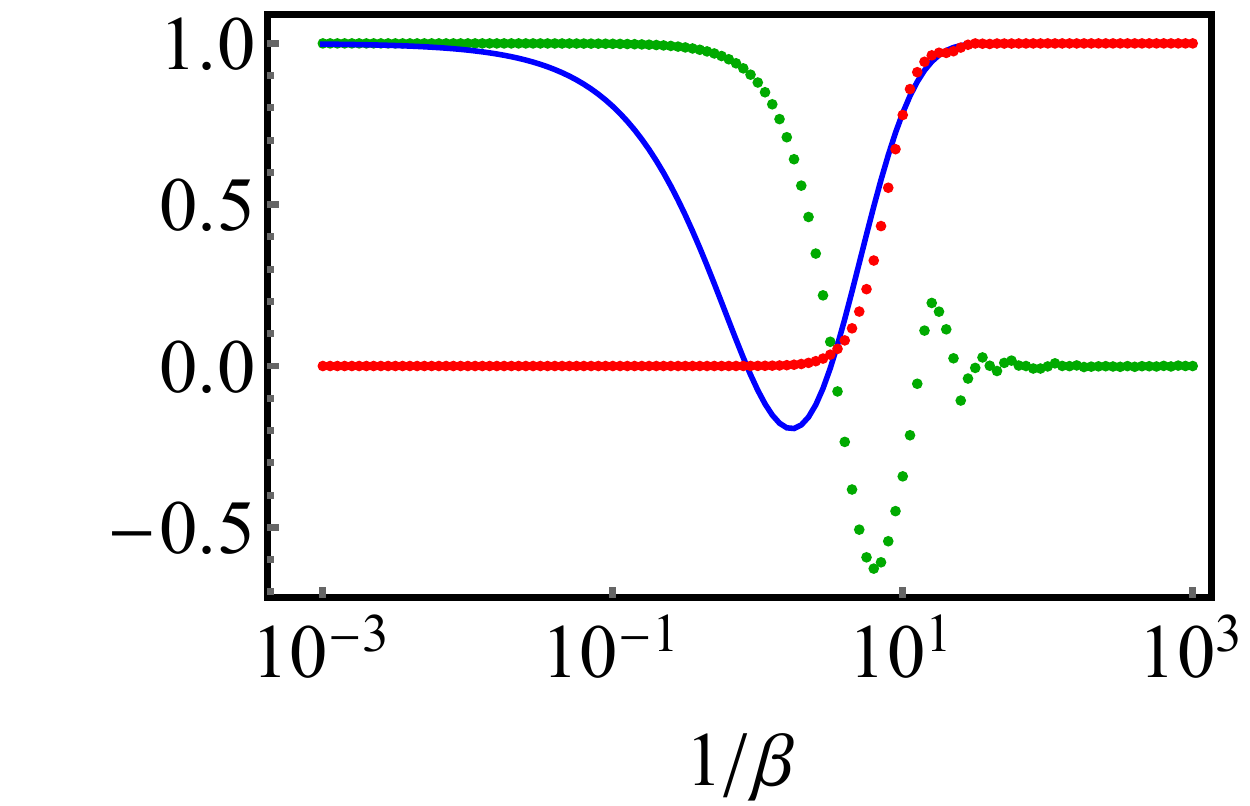}}
\caption{The numerically calculated $S_{1\rar 1}$ (green dots) and $S_{1\rar 5}$ (red dots) vs. $1/\beta$ for a $5$-site SSH chain \eqref{eq:5-SSH-Hamiltonian} under a linear quench from $t=1/(2\beta)$ to $t=1/(2\beta)$. For comparison, the blue curve is $S_{33}$ for a $t=-\infty$ to $t=\infty$ evolution of the model \eqref{eq:Schrodinger} with parameters \eqref{eq:Halpara-SSH}.}
\label{fig:5-SSH-edge}
\end{figure}

One may also be interested in a quench from $t=-1/(2\beta)$ to $t=1/(2\beta)$, as sketched in Fig.~\ref{fig:5-SSH}(c). Especially, let's take the system to be initially localized at the left edge $|1\ra$ and ask how this state will transfer under such a quench.  Let's use $S_{i\rar j}$ ($i,j=1,\ldots, 5$) to denote the scattering amplitude from a state $|i\ra$ to a state $|j\ra$ in the chain.  In Fig.~\ref{fig:5-SSH-edge} we plot numerical results of the scattering amplitudes $S_{1\rar 1}$ and $S_{1\rar 5}$ for this quench (both amplitudes turn out to be always real -- a result of symmetries). $S_{1\rar 1}$ corresponds to the probability that the state stays on the left end, and $S_{1\rar 5}$ the probability that the state transfers to the right end. We see that as $\beta$ decreases, $S_{1\rar 1}$ decreases from $1$ to $0$ after some oscillations, whereas $S_{1\rar 5}$ increases from $0$ to $1$. These tendencies are expected, since at large $\beta$ the state does not have time to transfer so $S_{1\rar 1}$ is close to $1$, whereas at small $\beta$ the evolution is adiabatic so $S_{1\rar 5}$ is close to $1$. Besides, we see that the amplitude $S_{1\rar 5}$ becomes close to the previously calculated $S_{33}$ (for a $t=-\infty$ to $t=\infty$ evolution of the model \eqref{eq:Schrodinger} with parameters \eqref{eq:Halpara-SSH}) at around $\beta<0.1$. Actually $S_{33}$ can be interpreted as the amplitude of adiabatic evolution to stay on the zero energy eigenstate; the left or right edge states at $t=\pm 1/(2\beta)$ are both instantaneous eigenstates along this adiabatic evolution. For $\beta$ small enough, the LZ transition happens well within the time interval from $t=-1/(2\beta)$ to $t=1/(2\beta)$, so the scattering amplitude $S_{1\rar 5}$ can be approximated by $S_{33}$ from a $t=-\infty$ to $t=\infty$ evolution.

\section{Conclusions and Discussions}

We study a class of 5-state LZ model for which integrability conditions are not helpful in finding its solutions, nor the other standard solving methods, either. 
We analyze the constraints on its scattering matrix and show that the transition probabilities depend on two independent unknowns. For particular choices of the Hamiltonian's parameters, by fitting to exact results from numerical simulations of the Schr\"{o}dinger equation we find that these two independent unknowns can be approximately expressed by analytical expressions. Thus, we obtain nearly exact solution of all transition probabilities of this model. For general parameter choices, we determine the transition probabilities in the diabatic limit up to leading orders of series expansions in terms of the inverse sweep rate. We also show that this model can describe a $5$-site SSH chain under a linear quench of its couplings.

We expect that the ``analytical constraint plus fitting'' method described in this work could be applied when dealing with other currently unsolved multistate LZ models. 
Generally, the method should work better for a model when the number of independent unknowns is small after using all analytical constraints on its scattering matrix. Thus, it would be interesting to look at other currently unsolved multistate LZ models with small numbers of levels and with symmetries as rich as possible.  Another possible future direction is to explore linear quenches in other systems besides SSH chains, and see if mappings to multistate LZ Hamiltonians in standard forms with relatively simple expressions of matrix elements exist. If so, by studying those multistate LZ models one can gain more understanding on the original systems.

Finally, we note that another interesting possibility is to perform a fitting with the help of  artificial intelligence (AI). The fitting problem we face here is basically to find an unknown symbolic expression to describe a set of data. Such kind of problem is known in computer science as {\it symbolic regression}, which is a rapidly growing research area \cite{Udrescu-2020,Udrescu-2020-2,La-Cava-2021}. In our case, the data set is generated from exact numerical solution of an equation (the Schr\"{o}dinger equation of the $5$-state LZ model) at different choices of parameters. This is exactly the type of problem studied in an interesting recent work by Ashhab \cite{Ashhab-2023}. In \cite{Ashhab-2023}, a state-of-the-art symbolic regression package based on machine learning, named AI Feynman, was run to look for symbolic solutions of three multistate LZ models: (a) the ($2$-state) LZ model, (b) a  LZ model coupled to a harmonic oscillator, and (c) a 3-state LZ model. For the first two models exact solutions  in terms of analytical expressions are known, whereas for the last model analytical expressions of only the probabilities to stay on the extremal levels are known according to the Brundolbler-Elser formula; the performance of the package can thus be tested against these exact solutions. The input data to the package also come from numerically solving the Schr\"{o}dinger equations. Thus, our problem on finding the transition probabilities of the 5-state model by fitting is basically the same as the ones described in \cite{Ashhab-2023} -- Ashhab let the AI package do exactly the same job that we did in this work by our own brains, namely, to guess some analytical expressions out of numerically obtained data. In \cite{Ashhab-2023}, for the model (a) the package was able to reproduce the exact analytical solution, namely, the LZ formula, but for the models (b) and (c) it reproduced only a small portion of the known analytical solutions. Thus, the package's current performance seems still not better than human brains'. But given the rapidly increasing abilities of AI nowadays, it is natural to expect that in the future AI will  outperform human brains  in such a job on finding analytical solutions, and thus it will  become a powerful tool in treating current unsolved multistate LZ  problems like the one considered in this paper, or even a more general quantum evolution problem. Future researches in this direction are definitely worthwhile.

\section*{Acknowledgements}
We are grateful for helpful discussions with Nikolai A. Sinitsyn. This work was supported by National Natural Science Foundation of China under No. 12105094 (R. H and C. S.) and under No. 12275075 (F. L.). We also thank the support by the Fundamental Research Funds for the Central Universities from China.

\section*{Appendix A: Simplification of constraints on the scattering matrix}

\setcounter{figure}{0}
\setcounter{equation}{0}
\renewcommand{\theequation}{A\arabic{equation}}
\renewcommand\thefigure{A\arabic{figure}}

In this appendix, we present details of simplification of the constraints on the elements of the scattering matrix $S$, namely, Eqs.~\eqref{eq:cons-bipartite}-\eqref{eq:cons-HC4}.

We first demonstrate that, although there are originally $6$ complex unknowns, the system of equations depends on only $4$ phases. There are $8$ equations that involve phases: the last two HCs \eqref{eq:cons-HC3} and \eqref{eq:cons-HC4}, and the last $6$ unitarity constraints \eqref{eq:unitary-4}-\eqref{eq:unitary-9}. In terms of the $4$ rotated elements defined in \eqref{eq:S21-tilde}-\eqref{eq:S42-tilde}, the $6$ unitarity constraints involving phases read:
\begin{align}\label{}
&2S_{11}\tilde{S}_{51}+2\tilde{S}_{21}\tilde{S}_{41}+|S_{31}|^2=0,\label{eq:unitary-tilde-1}\\
&2\tilde{S}_{21}^*\tilde{S}_{41}+2S_{22}\tilde{S}_{42}+|S_{32}|^2=0\label{eq:unitary-tilde-2},\\
&(S_{22}-S_{11})\tilde{S}_{21} +|S_{31}S_{32}|+\tilde{S}_{41}\tilde{S}_{42}^*-\tilde{S}_{51}\tilde{S}_{41}^*=0,\label{eq:unitary-tilde-3}\\
&(S_{33}-S_{11})|S_{31}|+\tilde{S}_{21}|S_{32}|+\tilde{S}_{41}|S_{32}|-\tilde{S}_{51}|S_{31}|=0,\label{eq:unitary-tilde-4}\\
&(S_{22}-S_{11})\tilde{S}_{41}+\tilde{S}_{21}\tilde{S}_{42}+|S_{31}S_{32}|-\tilde{S}_{51}\tilde{S}_{21}^*=0,\label{eq:unitary-tilde-5}\\
&(S_{22}+S_{33}) |S_{32}|+\tilde{S}_{21}^* |S_{31}|+\tilde{S}_{42}|S_{32}|+\tilde{S}_{41}|S_{31}|=0.\label{eq:unitary-tilde-6}
\end{align}
We see the phases of $ S_{31} $ and $ S_{32}$ drop out from all these equations. This also happens in the $3$rd HC \eqref{eq:cons-HC3}, which becomes:
\begin{align}\label{}
& Y S_{33}-S_{11}|S_{32}|^2+S_{22}|S_{31}|^2- |S_{31}S_{32}|(\tilde{S}_{21}+\tilde{S}^*_{21})=Y.\label{eq:cons-HC3-tilde}
\end{align}
The $4$th HC \eqref{eq:cons-HC4} involves calculation of the determinant of a $4\times 4$ matrix. After some tedious algebra with repetitive usages of other constraints, we find that it reduces to a simple form
\begin{align}\label{eq:cons-HC4-tilde}
&[(1+S_{11})^2-|S_{51}^2| ](S_{11}-S_{22})+ |S_{31}|^2[1+S_{11}-\frac{1}{2}(\tilde{S}_{51}+\tilde{S}^*_{51})]  =0.
\end{align}
Therefore, the system of equations involves only $4$ complex unknowns, or equivalently, $4$ phases.

We next show that all $4$ rotated elements (or equivalently, their $4$ phases) can be determined up to a sign by the real elements and the magnitudes of complex elements. From \eqref{eq:unitary-tilde-1} and \eqref{eq:unitary-tilde-2}, we get:
\begin{align}\label{}
&\tilde{S}_{51}=-\frac{1}{S_{11}}(\tilde{S}_{21}\tilde{S}_{41}+\frac{1}{2}|S_{31}|^2),\label{eq:51in2141-1} \\
&\tilde{S}_{42}=-\frac{1}{S_{22}}(\tilde{S}_{21}^*\tilde{S}_{41}+\frac{1}{2}|S_{32}|^2).\label{eq:42in2141}
\end{align}
On the other hand, from \eqref{eq:unitary-tilde-4}
, we get:
\begin{align}\label{}
&\tilde{S}_{51}=S_{33}-S_{11}+(\tilde{S}_{21}+\tilde{S}_{41})\frac{|S_{32}|}{|S_{31}|}.\label{eq:51in2141-2}
\end{align}
From Eqs.~\eqref{eq:51in2141-1} and \eqref{eq:51in2141-2}, we can express $\tilde{S}_{41}$ in terms of $\tilde{S}_{21}$ as:
\begin{align}\label{eq:41in21}
&\tilde{S}_{41}=-\frac{S_{11}(S_{33}-S_{11})+\frac{1}{2}|S_{31}|^2+S_{11}\tilde{S}_{21}\frac{|S_{32}|}{|S_{31}|}}{\tilde{S}_{21}+S_{11}\frac{|S_{32}|}{|S_{31}|}}.
\end{align}
Therefore, $\tilde{S}_{41}$, $\tilde{S}_{51}$, and $\tilde{S}_{42}$ can all be expressed in terms of the three real elements $S_{11}$, $S_{22}$ and $S_{33}$, the two magnitudes $|S_{31}|$ and $|S_{32}|$, and $\tilde{S}_{21}$. Besides, Eq.~\eqref{eq:cons-HC3-tilde} shows that $\tilde{S}_{21}+\tilde{S}_{21}^*$ can be expressed in terms of the three real elements and the two magnitudes. So if all the real elements and the magnitudes of complex elements are known, all the $4$ unknown phases can be determined up to a sign. There is still a sign ambiguity when determining the phases of $\tilde S_{21}$. Actually, each the original $14$ constraints is invariant if we send the $4$ rotated elements to their complex conjugates. Thus, the sign of the phase of $\tilde S_{21}$ cannot be determined by the real elements and the magnitudes. This sign needs to be fixed by the solution of the actual evolution process.

Finally, we show that all the real elements and the magnitudes of complex elements can be expressed in terms of two of them. There are totally $9$ of them -- $3$ real elements and $6$ magnitudes of complex elements. The original set of constraints already contain $6$ independent relations on them, namely, Eqs.~\eqref{eq:cons-bipartite}-\eqref{eq:unitary-3} and Eqs.~\eqref{eq:cons-HC1}-\eqref{eq:cons-HC2}. From the constraints involving phases, we can derive one more constraint on the real elements and the magnitudes. Plugging Eq.~\eqref{eq:41in21} into Eq.~\eqref{eq:51in2141-2}, we express $\tilde{S}_{51}$ in terms of $\tilde S_{21}$ as:
\begin{align}\label{}
&\tilde{S}_{51}=\frac{1}{2}(S_{33}+1)-X \frac{   X(1+S_{33})-|S_{32}|^2 }{2Y} +\frac{|S_{32}|}{|S_{31}|} \frac{    X(1+S_{33})-|S_{32}|^2+2Y   }{2Y}\tilde{S}_{21}.
\end{align}
Calculating the real part of $\tilde{S}_{51}$ with the help of some constraints involving only real elements and magnitudes, we get:
\begin{align}\label{}
&\frac{1}{2}(\tilde S_{51}+\tilde S_{51}^*)=\frac{1}{2}(S_{33}+1)+X  +\frac{S_{33}-1}{8Y|S_{31}|^2}[  X(1+S_{33})-|S_{32}|^2+2Y ]^2.
\end{align}
Plugging this into \eqref{eq:cons-HC4-tilde} and simplifying, we get an expression of $|S_{51}|^2$ in terms of $|S_{32}|$ and $S_{33}$ only:
\begin{align}\label{}
&|S_{51}|^2 
=\frac{1}{4}(1+S_{33})^2+ |S_{32}|^2 +X^2-Y- Z.
\end{align}
This equation and the original $6$ constraints  Eqs.~\eqref{eq:cons-bipartite}-\eqref{eq:unitary-3} and Eqs.~\eqref{eq:cons-HC1}-\eqref{eq:cons-HC2} form the set of  constraints on the real elements and the magnitudes. We checked that there are no other independent relations on the real elements and the magnitudes. Thus, the total number of independent equations on the real elements and the magnitudes is $7$, and we will have $9-7=2$ independent real elements and magnitudes after these $7$ equations are used. If we choose the two independent unknowns to be $S_{33}$ and $|S_{32}|$, the rest $7$ are then expressed as Eqs.~\eqref{eq:S11}-\eqref{eq:S42} in the main text.

Summarizing, from the $14$ constraints \eqref{eq:cons-bipartite}-\eqref{eq:cons-HC4} on the elements of the scattering matrix $S$, all unknowns can be determined in terms of two of them (e.g. $S_{33}$ and  $|S_{32}|$), up to an overall sign of the phases of the complex unknowns.

 \end{document}